\documentclass[conference]{IEEEtran}
\usepackage{color, soul}

\usepackage{multicol}

\usepackage[utf8]{inputenc}
\usepackage[T1]{fontenc}
\usepackage[british]{babel}

\ifCLASSINFOpdf
\else
\fi
\begin{document}
%
\title{A Survey on Essential Components of a Self-Sovereign Identity}

\author{
\IEEEauthorblockN{Alexander Mühle, Andreas Grüner, Tatiana Gayvoronskaya, Christoph Meinel}
\IEEEauthorblockA{Hasso Plattner Institute\\
Potsdam, Germany\\
\{alexander.muehle, andreas.gruener, tatiana.gayvoronskaya, christoph.meinel\}@hpi.de}
}


%


\maketitle

\begin{abstract}
This paper provides an overview of the Self-Sovereign Identity (SSI) concept, focusing on four different components that we identified as essential to the architecture. Self-Sovereign Identity is enabled by the new development of blockchain technology. Through the trustless, decentralised database that is provided by a blockchain, classic Identity Management registration processes can be replaced. \\
We start off by giving a simple overview of blockchain based SSI, introducing an architecture overview as well as relevant actors in such a system. We further distinguish two major approaches, namely the Identifier Registry Model and its extension the Claim Registry Model. \\
Subsequently we discuss identifiers in such a system, presenting past research in the area and current approaches in SSI in the context of Zooko's Triangle. As the user of an SSI has to be linked with his digital identifier we also discuss authentication solutions.\\
Most central to the concept of an SSI are the verifiable claims that are presented to relying parties. Resources in the field are only losely connected. We will provide a more coherent view of verifiable claims in regards to blockchain based SSI and clarify differences in the used terminology.\\
Storage solutions for the verifiable claims, both on- and off-chain, are presented with their advantages and disadvantages.\\
\end{abstract}


%
\IEEEpeerreviewmaketitle

\section{Introduction}
Blockchain technology has experienced tremendous hype in recent years and is touted as a transformative evolution in distributed systems \cite{meinel2018hype}. Satoshi Nakamoto is seen as father of the technology for introducing \textit{Bitcoin: A peer-to-peer electronic cash system} \cite{nakamoto2008bitcoin}. By applying the concept of trustless timestamping proposed by Haber and Stornetta \cite{haber1990time} to a decentralised setting and combining it with a chain of Proof-of-Work \cite{dwork1992pricing} \cite{back2002hashcash} the so called Nakamoto consensus protocol was established.\\
The computational resources invested in the Proof-of-Work solutions are equivalent to votes on the correct version of the blockchain, so as long as more than 50\% of the computational resources are in control of honest nodes, an eventual consistency can be achieved \cite{vukolic2016eventually}.\\
This decentralised consensus protocol has seen application in numerous fields, one of them identity mangement.\\
The management of identites has also experienced increased interest due to the ever growing need of digital identites, as a large part of peoples lifes is spent online, interacting with services. A digital identity can be simply described as a means for people to prove electronically that they are who they say they are and distinguish different entities from one another.\\
Although the term ``\textit{Self-Sovereign Identity}'' (SSI) is still only loosely defined, a few key properties of the concept have emerged. In essence it is an identity management system which allows individuals to fully own and manage their digital identity. The W3C working group on verifiable claims states that in a self-sovereign identity system users exist independently from services \cite{vctfFAQ}. This highlights the contrast to current identity management which either relies on a number of large identity providers such as Facebook (Facebook Connect) and Google (Google Sign-In) or the user has to create new digital identities at each individual service provider.\\
Christopher Allen proposed \textit{Ten Principles of Self-Sovereign Identity} \cite{allen2016path} which layed out the requirements for a system implementing the self-sovereign identity concept. 
These Principles were further grouped into the three categories \textit{security}, \textit{controllability}, and \textit{portability} in a whitepaper by the Sovrin Foundation \cite{tobin2016inevitable} as pictured in Figure \ref{fig:sovrin}.
\begin{figure}[!h]
\begin{center}
\begin{tabular}{l|c|r}
Security & Controllability & Portability\\
\hline
Protection & Existence & Interoperability\\
Persistance & Control & Transparency \\
Minimisation & Consent & Access
\end{tabular} 
\\[0.2cm]
\caption{Christopher Allen's Ten Principles of Self-Sovereign Identity summarised by the Sovrin Foundation \cite{tobin2016inevitable}}
\label{fig:sovrin}
\end{center}
\end{figure}
\\Essentially \textit{security} can be boiled down to the protection of personal user data and the limiting of data exposure to the minimum required to fulfill a function. Additionally a persistent identity was named as a security requirement. Persistence in this context however should not contradict a ``right to be forgotten'' according to Allen. 
This right to be forgotten could also be grouped into the \textit{controllability} category as both \textit{control} and \textit{consent} should extend to the removal of the identity not only the creation and access.\\
\begin{figure*}[!h]
\begin{picture}(400,135)
\put(220, 100){\framebox(60,30){User}}
\put(250, 100){\vector(0,-1){30}}
\put(255, 80){control}
\put(220, 40){\dashbox(60,30){Identity}}
\put(150, 55){\vector(1,0){70}}
\put(170,58){issues}
\put(90, 40){\framebox(60,30){Claim-Issuer}}
\put(280, 55){\vector(1,0){70}}
\put(300,58){presents}
\put(350, 40){\framebox(70,30){Relying Party}}
\put(240,10){trusts}
\qbezier(115,40)(225,10)(390,40)
\end{picture}
\caption{Self-Sovereign Identity Actors}
\label{fig:actors}
\end{figure*}
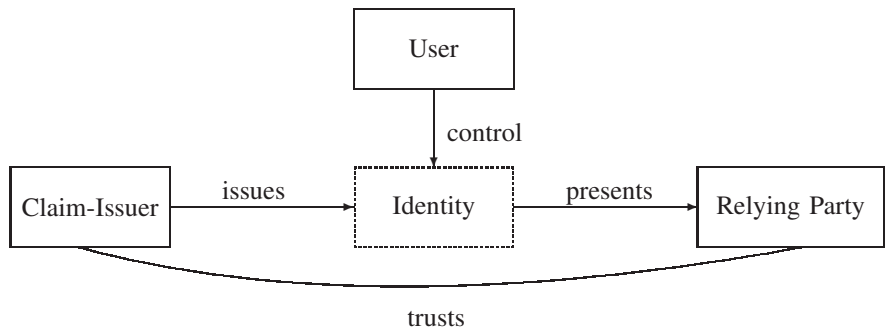
Another essential requirement for an SSI system is the portability of the identity. Allowing the user to use their identity wherever they want and being independent of any particular identity provider.\\
Although there are a large number of projects and initiatives concentrated on Self-Sovereign Identity, both the terminology and understanding of architectures differs widely.\\
New innovations come primarely from private ventures or individuals volunteering in working groups. While this leads to a lot of interest in the wider public, the documentation of such ideas is either very practical or only for advertisement purposes limiting their scientific usefulness.\\
This paper's objective is to give an overview and deeper understanding of the concept of SSI as well as the current state of the art. For this purpose we will look at four basic components needed in a Self-Sovereign Identity system.\\
Before we start going into detail about the different components we will first provide a high level overview of the Self-Sovereign Identity (SSI) Architecture. After this general overview in Section \ref{sec:overview} we will present the essential components of such a system, starting with the identifier to be chosen for identities in the system in Section \ref{sec:identifiers}. Further the authentication of the identity will be discussed in Section \ref{sec:authentication}. The concept of verifiable claims and their integral role in the SSI system as well as the possibility for reputation systems will be reviewed in Section \ref{sec:reputation}. For privacy and scalability considerations we will also discuss storage approaches for use in a Self-Sovereign Identity system in Section \ref{storage}.\\


%
\section{Self-Sovereign Identity Architecture}
\label{sec:overview}
\begin{figure*}[!h]
\begin{picture}(400,280)
\put(190,200){\dashbox{0.5}(115, 75)}
\put(190,265){\dashbox{0.5}(40, 10){Storage}}
\put(205,210){\framebox(85,40)}
\put(205,252){Verifiable Claim (VC)}
\put(235,235){Claim}
\put(210, 232){\line(6,0){75}}
\put(225,220){Attestations}
\put(250, 170){\line(0,6){40}}
\put(255, 180){Control}
\put(20,140){\framebox(100,30){Issuer}}
\put(120,155){\line(6,0){80}}
\put(135,160){Issue VC}
\put(200,140){\framebox(100,30){User-Agent}}
\put(300,155){\line(6,0){80}}
\put(310,160){Present VC}
\put(380,140){\framebox(100,30){Verifier}}
\put(130,30){\dashbox{0.5}(240, 75)}
\put(130,95){\dashbox{0.5}(50, 10){Blockchain}}
\put(151, 78){Identifier Registry}
\put(150,45){\framebox(90,30){}}
\put(155, 55){Identifier $\mapsto$ Auth}
\put(275, 78){Claims Registry}
\put(260,45){\framebox(90,30){}}
\put(265, 55){Hash(VC)}
\put(230, 75){\line(0,6){65}}
\put(145,120){Register Identifier}
\put(270, 75){\line(0,6){65}}
\put(285,120){Register VC}
\end{picture}
\caption{Self-Sovereign Identity Architecture}
\label{fig:architecture}
\end{figure*}
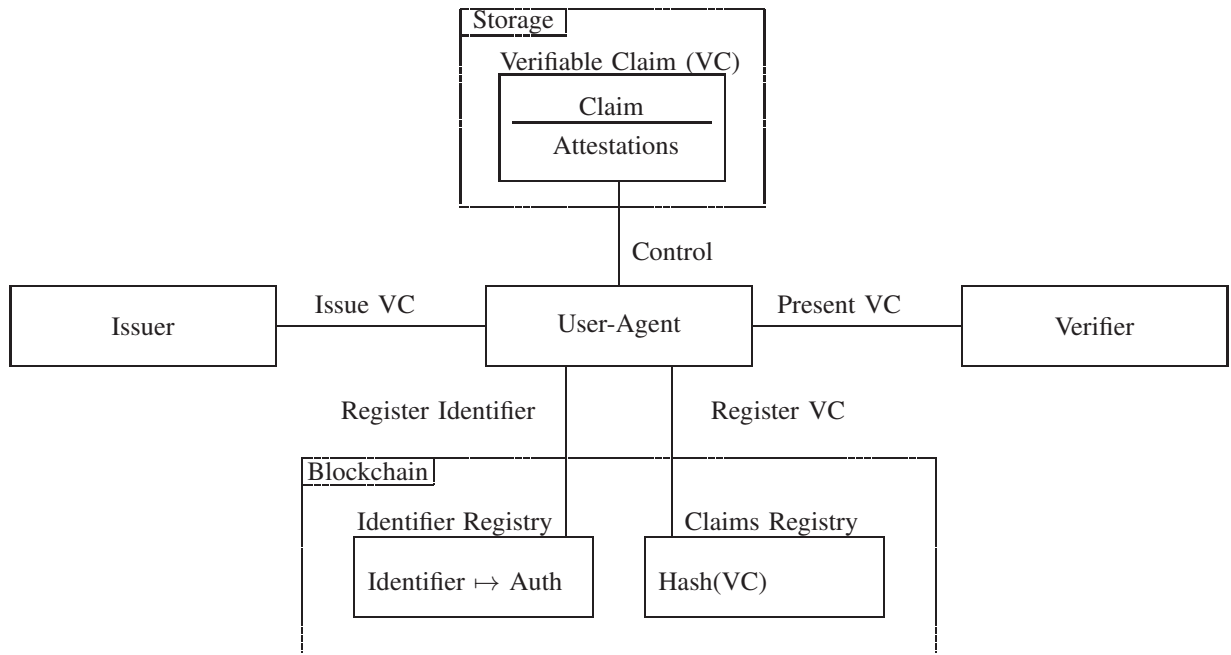
In contrast to most previous identity management systems where the service provider was at the center of the identity model, SSI is user centric. In Figure \ref{fig:actors} the relation between the different actors of the system can be observed. The claim issuer issues (at least part of) the identity by attesting to certain attributes of the user. This identity is controlled by the user himself. Any relying party that needs to identify the user will be presented with the parts of the user controlled identity relevant to him. In order to accept the identity, the relying party needs to have a trustful relationship with the claim issuer.\\
The basis of this new architecture type is the distributed ledger of the blockchain. In Figure \ref{fig:architecture} the relation between the different components of a typical SSI architecture are layed out. The blockchain acts as replacement for the registration authority in classic identity mangement systems. In this paper we will call this blockchain function the \textit{identifier registry}. Here the pairing of \textbf{identification} and \textbf{authentication} is maintained. The identifier as well as the \textbf{verifiable claims} are directly managed by the user.\\
The identifier is tied to the specific user by use of an authentication method such as asymmetric cryptography. By establishing a pairing of identifier and public key on the blockchain the identifier can be verified by anyone reading the blockchain by posing a challenge to the user himself or a delegate of the user.\\ 
A distinction can be made between subject and holder in some cases, i.e guardians to underaged individuals or attorney client relations. In the following we will, for simplicity, assume that the holder is indeed the subject of any claims and will refer to him as user.\\
The actual identity claim is stored in the user controlled \textbf{storage}, typically off-chain for privacy considerations. The relying party, also called claim-verifier, can then compare the publicely available identifier with the identifier in the claim that has been handed to him by the user. After authenticating the user with the authentication method presented in the public blockchain, the claim itself can be verified and accepted or rejected by the relying party.\\
We will describe this architecture as \textit{Identifier Registry Model}. A very popular competing model can be described as \textit{Claim Registry Model}. In that model the blockchain not only functions as a registry for the identifiers of an identity but also to hold the cryptographic fingerprints of all the associated claims of an identity. This model can be seen as an extension to the Identifier Registry Model.\\
In this process no information about the user has to be stored at either the issuer or the verifier. Only the trust between the issuer and verifier has to be established beforehand.
As described in this section, the SSI architecture relies on the mapping of an identifier to a specific authentication method that is recorded on the blockchain. In the next two sections we will discuss how this identifier and its namespace is chosen as well as the authentication methods used.\\

\section{Identification}
\label{sec:identifiers}
Bryce Wilcox-O'Hearn published a widely cited article on namespaces in computer systems in 2001. In it he layed out what is now known as Zooko's Triangle \cite{wilcox2003names}. According to his assessement it was impossible (or highly unlikely) that someone would be able to design a system in which identifiers could be chosen in a distributed fashion but at the same time being both secure and human-readable.
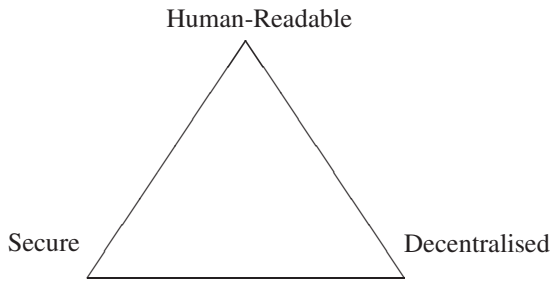
\begin{figure}[!h]
\begin{picture}(100,100)
\put(50,0){\line(2,3){60}}
\put(20,10){Secure}
\put(170,0){\line(-2,3){60}}
\put(80,95){Human-Readable}
\put(50,0){\line(6,0){120}}
\put(170,10){Decentralised}
\end{picture}
\caption{Zooko's Triangle}
\label{fig:zooko}
\end{figure}
\\
In this context distributed means without the need for a central registration and verification process, while secure refers to the identifiers being securely unique (collision free).\\
The identifiers that are presented in this section can be grouped into three different categories. Firstly the identifiers based on random number generation that rely on probabilities to avoid collisions. Secondly the centralised identifiers that utilise a registration authority in order to assign identifiers and prevent collisions. Finally we will discuss how the blockchain technology can help merge the best aspects of both these approaches. \\
Already in the 1980s the need for a globally unique identifier became apparent \cite{levine1987apollo}. The Universally Unique Identifier (UUID or GUID) \cite{rfc4122} does not require a central registration process but rather lets users generate their own identifiers, therefore partially fulfilling the decentralisation requirement formulated by Zooko's Triangle. In the UUID versions 1 and 2 the uniqueness is guaranteed by inlcuding node specific information such as the users MAC address which are, unless tampered with, uniquely assigned by the manufacturer of the network card and the IEEE registration authority \cite{regauth}. This means in those versions there is still a centralised compononent while version 4 is using large random or pseudo-random numbers to avoid collisions. There are a total of $2^{122}$ possible version 4 UUIDs making a collision, assuming no implementation errors, highly unlikely to the point that it can be ignored \cite{jesus2006id}. This means UUIDs can be considered secure in this context and do not require a central registration authority. Apart from being non-human readable in the sense that no human could realisticly remember specific UUIDs, they are also not completely decentralised as the verification process of key-value pairs using UUIDs typically requires a trusted third party for verification.\\
This is where public/private key pairs hold a significant advantage over UUIDs as identifiers in an SSI. In contrast to a UUID a public/private key pair would not require a trusted third party for verification as it is self-authenticating.\\
A distinction between self-authenticating and non-self-authenticating key-value pairs helps in understanding Zooko's argument. Self-authenticating schemes such as secure hash algorithms or public/private keypairs can create key-value pairs collision free (to current knowledge) and verify them without third party input but are typically non-human readable identifiers.\\
In non-self-authenticating schemes however there needs to be trust placed in a third party, assigning and verifying the name-value pairs.\\
Although the public-private key pair used in a X.509 certificate is self-authenticating, the mapping of a human readable \textit{Distinguished Name} to a specific public key is not. For this mapping a centralised \textit{Certificate Authority} has to be trusted to correctly assign and store the pairing of name and public key. This centralisation carries significant risk though. Either through attacks \cite{prins2011diginotar} or coercion \cite{soghoian2011certified} the central authority can be compromised.\\
However, even more decentralised solutions such as PGP \cite{rfc4880} that do not rely on central entities for verification defacto utilise a quasi-centralised approach to assign human readable identifiers by using email-addresses. These are issued by a number of centralised providers that ultimately rely on ICANN to assign domain names without collisions.\\
%
Up until this point, Zooko's Triangle hypothesis held up. Either identifiers were not human readable or part of the decentralisation requirement was not fulfilled.
From 2011 on a number of name services on the blockchain appeared, ``squaring'' Zooko's Triangle. With the distributed ledger technology it is possible to choose a human-readable identifier in a decentralised fashion as well as assign and verify name-value pairs without third party input.\\
In contrast to previous decentralised human-readable namespaces (i.e. as initially used in Freenet \cite{clarke2001freenet}) that were unsafe, the consensus protocol of the blockchain and the global view of the system can guarantee that once a name-value pair has been established it can not be changed without the correct authentication and most importantly the same identifier can't be assigned more than once. As there is no central authority assigning name-value pairs however, there need to be other mechanisms.\\
The first name service built on Bitcoin called Namecoin \cite{namecoin} as well as a later competitor Emercoin \cite{emercoin} used first come first serve logic to assign name-value pairs. This policy however causes problems such as squatting of names, which was further escalated by the lack of centralised control. Kalodner et al. found in their study of the Namecoin namespace design that of the 120,000 registered domain names, only 28 were not squatted or had non-trivial content \cite{kalodner2015empirical}.
They argue that because the names are human readable they are naturally scarce and will therefore hold some market value compared to the essentially infinite non-human readable identifiers such as hashes of keys or the public key to a private key. \\
Carl Ellison stated in his 1996 paper on \textit{Establishing Identity Without Certification Authorities} that ``\textit{it is clear that there is no such thing as a universal, global name space with names meaningful to all possible users and that there never will be}'' \cite{ellison1996establishing}. Ellison reasoned that there are simply too many names for a human to remember and attach meaning to.\\
These assessements were utilised by the Ethereum Nameservice (ENS) \cite{ens} which implemented a decentralised bidding process to reduce the problem of squatters.\\ 
The Self-Sovereign Identity system uPort \cite{uPort} uses an Ethereum smart contract address as persistent identifier for a users identity. The address is derived from the public key of the creator of the smart contract. Since this identifier is non-human readable, Christian Lundkvist of uPort sees ENS as a viable naming layer to map the non-human readable uPort ID to a human readable address \cite{lundkvist2017reddit}. Blockstack \cite{ali2017blockstack} similarely uses its blockchain Name System to implement a naming service with human readable names that a Blockstack identity can be linked to for their system.\\
W3C decentralised identifiers \cite{w3c2018did} can be seen as an even higher level naming scheme, similar to URNs. They resulted from an effort by a number of working groups investigating decentralised name systems. A decentralised identifier (DID) is comprised of a scheme as well as a method and method specific identifier. The method closely resembles the namespace component of an URN. Each distinct blockchain or rather each identity registry (there can be multiple per blockchain, i.e uPort, Civic \cite{civic}, SelfKey \cite{selfkey} all operate on Ethereum) constitutes its own namespace while the blockchain specific identifiers such as a uPort ID or ENS name specify the actual identity addressed by the DID. An example for such a DID path would be: did:examplechain:123456789\\
The Decentralized Identity Foundation is developing a universal resolver for these DID paths. Currently Sovrin \cite{sovrin}, Bitcoin \cite{nakamoto2008bitcoin}, Blockstack \cite{ali2017blockstack}, uPort \cite{uPort}, Interplanetary Identifiers \cite{ipid}, and Veres One \cite{veresone} are supported by the resolver with implemented drivers. The resolver uses the method type to decide which driver to use and uses the method specific identifier to resolve to the DID document stored on the specified Blockchain. The DID document or its equivalent in other systems is the key to the decentralised identity.\\
In it the authentication method is defined to bind the specified identifier to an identity that is in control of a secret key or other data used in the authentication. 

\section{Authentication}
\label{sec:authentication}
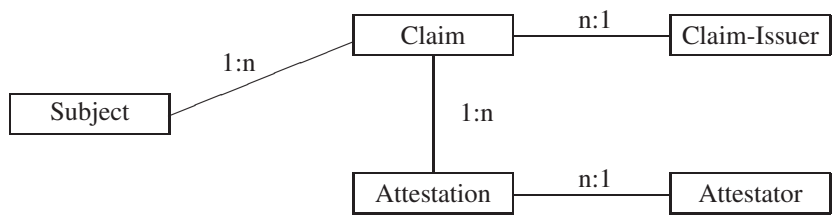
\begin{figure*}[!h]
\label{fig:vc}
\begin{picture}(400,100)
\put(100,50){\framebox(60,15){Subject}}
\put(160, 57){\line(5,2){70}}
\put(180, 74){1:n}
\put(230,80){\framebox(60,15){Claim}}
\put(290, 87){\line(6,0){60}}
\put(315, 90){n:1}
\put(350,80){\framebox(60,15){Claim-Issuer}}
\put(260, 35){\line(0,6){45}}
\put(270, 55){1:n}
\put(230,20){\framebox(60,15){Attestation}}
\put(290, 27){\line(6,0){60}}
\put(315, 30){n:1}
\put(350,20){\framebox(60,15){Attestator}}
\end{picture}
\caption{Relation between components in a verifiable claim}
\end{figure*}
In a Self-Sovereign Identity system authentication is typically done with the use of a public/private key pair where the public key is stored as value of the identifier on the blockchain. This concept has been described as \textit{Decentralised Public Key Infrastructure} \cite{fromknecht2014decentralized} \cite{allen2015dpki}. Thanks to the zero knowledge proof properties of the asymmetric cryptography it is possible to prove that a given user is indeed in control of the identity with the public key stored on the blockchain. Most popular Self-Sovereign Identity systems use a asymmetric cryptography authentication protocol.\\
This poses the question of how the user should hold the private key associated with his key pair. Blockstack uses probably the simplest solution where the keys are stored with the device that the identity was created on and the user himself is responsible for key recovery and mobility. To make this process somewhat more usable mnemonic phrases, typically of 12 words, are used as seed to generate the keys. Using those phrases it is possible to recreate the private key, reducing the effort needed to move keys from one system to the other.\\
The most used solution in the space currently however is utilising smartphones for key storage. This has the advantage of being more portable than other solutions. The challenge that is being posed by the relying party is communicated to the smartphone via a QR code displayed on the login page and the response directly sent from the smartphone to the designated endpoint. This visual communication removes the need for physical connections and therefore hardware support that would be needed for alternative mobile solutions such as SmartCards. David Chadwick already stated in 1999 that ``\textit{smart cards are beneficial in some scenarios [...] in some user environments, the costs and inconveniences clearly outweigh the potential benefits of using smart cards}'' \cite{chadwick1999smart}. Especially the need to equip workstations with card readers was seen as a major hinderence. \\
This is however not the only way authentication on the blockchain can be realised. Buldas et al. from Guardtime proposed a hash sequence authentication method for use in their blockchain system \cite{buldas2014efficient}. Their aim was to make their infrastructure more quantum computing resistant under the assumption that hash functions would stay secure in a quantum computing environment. The concept of hash sequence authentication has first been proposed by Lamport in 1981 \cite{lamport1981password}.\\
Another authentication method that has seen interest is the use of biometric systems. However most biometric cryptosystems need biometric dependent information (helper data) which could potentially reveal significant information about the original biometric template \cite{rathgeb2011survey}.\\
The W3C DID working group proposes the use of external biometric services in combination with a cryptographic hash of the biometric templates. \\
In theory any authentication method could be used through an identification service endpoint as defined in the W3C Verifiable Claims Working Group specification draft \cite{w3c2018draft}, however a self-authenticating method such as public key cryptography or hash sequences do not need to rely on any third party endpoints, eliminating yet another point of centralisation.\\
As the authentication in such a case relies on a secret held by the end-user it would be beneficial to provide him with a way for key recovery/replacement. In the DID scheme this is done by seperating authentication from authorisation allowing others to also change the DID document, i.e. changing the authentication key after the private key was lost. \\
uPort uses a quorum based key recovery where the holder logic includes a way for previously selected delegates to vote on replacing the public/private key pair of the user.\\
Key recovery seems to be a necessity for a working SSI system, since key losses are inevitable as the experience from bitcoin and other cryptocurrencies shows. In Bitcoin's case up to a quarter of all current coins have been lost due to unrecoverable private keys \cite{forbes2017lost}.\\

\section{Verifiable Claims}
\label{sec:reputation}
At the center of the Self-Sovereign Identity concept lay the \textit{verifiable claims}. The first clarification that is neccesary in this context is between a \textit{claim} and a verifiable claim. A claim in itself is only a statement about a specific subject. A \textit{credential}, which some differentiate from claims \cite{w3c2018issue}, describes a number of claims together with their meta data such as issuer and validity period.\\
Verifiable claims are verifiable through a signature of an attestation issuer that has either issued the claim himself or can attest the correctness of it. An \textit{attestation} can be seen as a proof in form of a signature attesting to a certain claim and meta data needed for verification such as name, validity period and signature scheme.\\
The verifiable claims themselves have to be associated with a subject, typically by including the subject identifier. This can be observed in Figure \ref{fig:vc} where the relation of different components and actors in a verifiable claim is shown. In addition to the subject, a verifiable claim should hold information about one or more actual claims as well as some meta data. The claim is issued by exactly one claim issuer. Similar to X.509 certificates meta data in a verifiable claim could include a validity period, the identity of the issuer and algorithms used for signature/encryption. To make the claim verifiable and trustable, the issuer has to sign the claim with a well known key. This is shown in Figure \ref{fig:vc} where each claim can have multiple attestations and each attestation has one attestator.\\
There are mainly two different ways for claims and attestations to be linked to a users identity. uPort which operates on the Ethereum network utilises smart contracts to keep a registry for claims on the blockchain. In the registry they maintain a mapping of user to hashes of claims that are stored off-chain. Through this fingerprint the integrity of a claim can be verified by relying parties. More specifially the timestamping property of the blockchain is utilised to prevent secret modification of a claim and its signature. This architecture however only allows the user to add new claims to his identifier unless a more complex access management is implemented for the uPort registry. A registry model inspired by the uPort registy is being standardised in the Ethereum Improvement Proposal (EIP) ERC780 \cite{erc780}. In the proposed Ethereum Claims Registry the writing of claims is not limited to the owner of the identity but issuers can directly add new claims and also revoke them in the registry.\\
The W3C VCWG data model on the other hand does not utilise a claim registry. They only rely on the blockchain for the mapping between an identifier and an authentication method. By including the identifier in the claim and having the issuer sign it, already secures against tampering from outside sources, however not against tampering by the issuer of whoever holds the signing key of the issuer. When colluding with the holder of the verifiable claim changes to the claim would go undetected and backdating or similar attacks could be done.\\
However, W3C's approach is very privacy preserving in the fact that not even the existence of new claims can be derived from blockchain changes but it also doesn't leverage the blockchains ability to trustfully timestamp items. Claims that have been altered after first issuance would need to be updated in the uPort registry, which would be recorded on the blockchain, protecting against tampering by the claims issuer or anyone in control of the singing key.\\
Another advantage that the registry model provides is the ease of revocation. As there is a ``central'' (but physically decentralised) location for all claims it would be possible to extend the registry with a revocation mechanism.\\
In comparison to current service centric/centralised identity solutions a relying party does not have to only trust a single issuance of a claim. Rather through the aggregation of multiple attestations for a claim, a more overarching and more decentralised trust model can be formed. This allows for relying parties to employ their own local confidence in certain attestators, depending on their individual relation. Working systems such as uPort which have their first real world applications \cite{zug} \cite{olympics}, although designed to support such reputation aggregation, so far only utilise the SSI as a way for single attestation claim verification in practice.\\
On a slightly higher level there is also the PGP like aggregation of multiple claims, not only attestations, which can be used to form a reputation model for an identity.

%
\section{Storage}
\label{storage}
\subsection{Public}
Although most data in SSI is stored off-chain some of the data is essential to have on-chain. Most importantly the already mentioned authentication such as a public key is typically included in a public fashion. In the end it is up to the user's disgression to decide what information he wants to publicly reveal and what he wants to control more closely. Both Blockstack and uPort have public profiles which not only include signing keys but also names and profile pictures. Especially Blockstack provides use-cases for public disclosure of information. Specifically social media accounts or PGP keys that need to be publicly available to realise their full potential are data that can be securely stored on the blockchain.
\subsection{Private}
In a lot of cases a user does not want to disclose claims about himself though. In most cases the privacy of the user has to be preserved. For this purpose most claims are stored off-chain not publicly available and either secured by the previously discussed claim registry model or simply linked by the identifier defined in the identifier registry.\\
Just as the public disclose of information, the user is also in control of where to store the claims. The most trustless way would be in a directly user controlled environment such as hardware in possession of the user. One such example would be the SelfKey project which utilises a users smartphone to store claims. This however poses some serious problems too. Namely data security both against data loss and data theft. The lack of data redundancy when locally storing claims on mobile devices as well as the security of the device itself have to be taken into consideration.\\ Blockstack on the other hand opted for centralised storage providers such as Amazon S3, Dropbox and Google Drive \cite{ali2017blockstack}. This helps prevent potential data loss as these systems are highly redundant. To minimise the impact of attacks on these systems, they are used in conjunction with each other, spreading the claim data over multiple providers.\\
Through the use of decentralised storage systems such as IPFS \cite{benet2014ipfs} uPort wants to minimise the reliance on centralised entities even more. IPFS is a peer-to-peer distributed file system based on distributed hash table technology and is only one example of decentralised storage a user can utilise.\\

\section{Future Work}
As we have hinted at in Section \ref{sec:reputation}, we consider the possibility for a reputation system for each individual claim an interesting future topic. Through the aggregation of multiple attestations, as well as weighting of different attestations a more complex than binary claim reputation might be realised.
In the same vain as a reputation model for a verifiable claim, the reputation for the identity as a whole could be derived from all verifiable claims associated with a given identity.

\section{Conclusion}
In the age of increasing digital interactions and analysis of user data, the concept of Self-Sovereign Identies has gained a large amount of interest. It promises its users more control and a more user-centric experience that, in contrast to previous user-centric efforts, does not have to rely on any centralised entities. The concept of verifiable claims has been extended by the Identity Registry Model as well as the Claim Registry Model. These decentralised registries were enabled by blockchain technology and altough not a necessasity the storage can be decentralised too. This only leaves the claim-issuers and their position of trust as centralised entities in the system.\\
In this paper the architecture of Self-Sovereign Identity systems has been presented as well as terms further clarified. Most importantly an analysis of essential components of such a system was provided.





\bibliographystyle{IEEEtran}
\bibliography{sources}

\begin{thebibliography}{10}
\providecommand{\url}[1]{#1}
\csname url@samestyle\endcsname
\providecommand{\newblock}{\relax}
\providecommand{\bibinfo}[2]{#2}
\providecommand{\BIBentrySTDinterwordspacing}{\spaceskip=0pt\relax}
\providecommand{\BIBentryALTinterwordstretchfactor}{4}
\providecommand{\BIBentryALTinterwordspacing}{\spaceskip=\fontdimen2\font plus
\BIBentryALTinterwordstretchfactor\fontdimen3\font minus
  \fontdimen4\font\relax}
\providecommand{\BIBforeignlanguage}[2]{{%
\expandafter\ifx\csname l@#1\endcsname\relax
\typeout{** WARNING: IEEEtran.bst: No hyphenation pattern has been}%
\typeout{** loaded for the language `#1'. Using the pattern for}%
\typeout{** the default language instead.}%
\else
\language=\csname l@#1\endcsname
\fi
#2}}
\providecommand{\BIBdecl}{\relax}
\BIBdecl

\bibitem{meinel2018hype}
Meinel, Gayvoronskaya, and Schnjakin, ``Blockchain: Hype oder innovation,''
  2018.

\bibitem{nakamoto2008bitcoin}
S.~Nakamoto, ``Bitcoin: A peer-to-peer electronic cash system,'' 2008.

\bibitem{haber1990time}
S.~Haber and W.~S. Stornetta, ``How to time-stamp a digital document,'' in
  \emph{Conference on the Theory and Application of Cryptography}.\hskip 1em
  plus 0.5em minus 0.4em\relax Springer, 1990, pp. 437--455.

\bibitem{dwork1992pricing}
C.~Dwork and M.~Naor, ``Pricing via processing or combatting junk mail,'' in
  \emph{Annual International Cryptology Conference}.\hskip 1em plus 0.5em minus
  0.4em\relax Springer, 1992, pp. 139--147.

\bibitem{back2002hashcash}
A.~Back \emph{et~al.}, ``Hashcash-a denial of service counter-measure,'' 2002.

\bibitem{vukolic2016eventually}
M.~Vukolic, ``Eventually returning to strong consistency.'' \emph{IEEE Data
  Eng. Bull.}, vol.~39, no.~1, pp. 39--44, 2016.

\bibitem{vctfFAQ}
Verifiable claims working group frequently asked questions.
  \url{http://w3c.github.io/webpayments-ig/VCTF/charter/faq.html}. Accessed:
  31/1/2018.

\bibitem{allen2016path}
C.~Allen. (2016) The path to self-sovereign identity.
  \url{http://www.lifewithalacrity.com/2016/04/the-path-to-self-soverereign-identity.html}.
  Accessed: 31/1/2018.

\bibitem{tobin2016inevitable}
A.~Tobin and D.~Reed, ``The inevitable rise of self-sovereign identity,''
  \emph{The Sovrin Foundation}, 2016.

\bibitem{wilcox2003names}
Z.~Wilcox-O’Hearn, ``Names: Decentralized, secure, human-meaningful: Choose
  two,'' 2003.

\bibitem{levine1987apollo}
P.~H. Levine, ``The apollo domain distributed file system,'' in
  \emph{Distributed Operating Systems}.\hskip 1em plus 0.5em minus 0.4em\relax
  Springer, 1987, pp. 241--260.

\bibitem{rfc4122}
Leach, Mealling, and Salz, ``A universally unique identifier (uuid) urn
  namespace,'' \emph{RFC4122}.

\bibitem{regauth}
{IEEE Standards Association}. Registration authority.
  \url{http://standards.ieee.org/develop/regauth/}. Accessed: 31/1/2018.

\bibitem{jesus2006id}
P.~Jesus, C.~Baquero, and P.~S. Almeida, ``Id generation in mobile
  environments,'' 2006.

\bibitem{prins2011diginotar}
J.~R. Prins and B.~U. Cybercrime, ``Diginotar certificate authority
  breach’operation black tulip’,'' \emph{Fox-IT, November}, 2011.

\bibitem{soghoian2011certified}
C.~Soghoian and S.~Stamm, ``Certified lies: Detecting and defeating government
  interception attacks against ssl (short paper),'' in \emph{International
  Conference on Financial Cryptography and Data Security}.\hskip 1em plus 0.5em
  minus 0.4em\relax Springer, 2011, pp. 250--259.

\bibitem{rfc4880}
Callas, Donnerhacke, Finney, Shaw, and Thayer, ``Openpgp message format,''
  \emph{RFC4880}.

\bibitem{clarke2001freenet}
I.~Clarke, O.~Sandberg, B.~Wiley, and T.~W. Hong, ``Freenet: A distributed
  anonymous information storage and retrieval system,'' in \emph{Designing
  privacy enhancing technologies}.\hskip 1em plus 0.5em minus 0.4em\relax
  Springer, 2001, pp. 46--66.

\bibitem{namecoin}
Namecoin. \url{https://namecoin.org}. Accessed: 29/1/2018.

\bibitem{emercoin}
Emercoin. \url{https://emercoin.com}. Accessed: 29/1/2018.

\bibitem{kalodner2015empirical}
H.~A. Kalodner, M.~Carlsten, P.~Ellenbogen, J.~Bonneau, and A.~Narayanan, ``An
  empirical study of namecoin and lessons for decentralized namespace design.''
  in \emph{WEIS}, 2015.

\bibitem{ellison1996establishing}
C.~Ellison \emph{et~al.}, ``Establishing identity without certification
  authorities,'' in \emph{USENIX Security Symposium}, 1996, pp. 67--76.

\bibitem{ens}
{Ethereum Name Service}. \url{https://ens.domains}. Accessed: 29/1/2018.

\bibitem{uPort}
Consensys. \url{https://uport.me}. Accessed: 14/2/2018.

\bibitem{lundkvist2017reddit}
C.~Lundkvist. \url{https://www.reddit.com/r/ethereum/comments/5wi6cl/
  what\_is\_a\_uport\_identity/deaki14/}. Accessed: 4/2/2018.

\bibitem{ali2017blockstack}
M.~Ali, R.~Shea, J.~Nelson, and M.~J. Freedman, ``Blockstack: A new
  decentralized internet,'' \emph{Whitepaper, May}, 2017.

\bibitem{w3c2018did}
{Drummond Reed, Manu Sporny, Dave Longley, Christopher Allen, Ryan Grant,
  Markus Sabadello}, ``Decentralized identifiers (dids) v0.7: Data model and
  syntaxes for decentralized identifiers,'' W3C, Tech. Rep., 01 2018.

\bibitem{civic}
C.~Technologies. \url{https://civic.com}. Accessed: 14/2/2018.

\bibitem{selfkey}
S.~Network. \url{https://selfkey.org}. Accessed: 14/2/2018.

\bibitem{sovrin}
S.~Foundation. \url{https://sovrin.org}. Accessed: 14/2/2018.

\bibitem{ipid}
J.~Holt. \url{https://github.com/jonnycrunch/ipid}. Accessed: 14/2/2018.

\bibitem{veresone}
D.~Bazaar. \url{https://veres.one}. Accessed: 14/2/2018.

\bibitem{fromknecht2014decentralized}
C.~Fromknecht, D.~Velicanu, and S.~Yakoubov, ``A decentralized public key
  infrastructure with identity retention.'' 2014.

\bibitem{allen2015dpki}
{Allen, Brock, Buterin, Callas, Dorje, Lundkvist, Kravchenko, Nelson, Reed,
  Sabadello, Slepak, Thorp, Wood}, ``Decentralized public key infrastructure: A
  white paper from rebooting the web of trust,'' \emph{Rebooting the Web of
  Trust}, 2015.

\bibitem{chadwick1999smart}
D.~Chadwick, ``Smart cards aren't always the smart choice,'' \emph{Computer},
  vol.~32, no.~12, pp. 142--143, 1999.

\bibitem{buldas2014efficient}
A.~Buldas, R.~Laanoja, and A.~Truu, ``Efficient quantum-immune keyless
  signatures with identity.'' \emph{IACR Cryptology ePrint Archive}, vol. 2014,
  p. 321, 2014.

\bibitem{lamport1981password}
L.~Lamport, ``Password authentication with insecure communication,''
  \emph{Communications of the ACM}, vol.~24, no.~11, pp. 770--772, 1981.

\bibitem{rathgeb2011survey}
C.~Rathgeb and A.~Uhl, ``A survey on biometric cryptosystems and cancelable
  biometrics,'' \emph{EURASIP Journal on Information Security}, vol. 2011,
  no.~1, p.~3, 2011.

\bibitem{w3c2018draft}
D.~Burnett, M.~Sporny, D.~Longley, and G.~Kellogg.
  \url{https://www.w3.org/TR/2017/WD-verifiable-claims-data-model-20170803/}.
  Accessed: 28/2/2018.

\bibitem{forbes2017lost}
J.~J. Roberts and N.~Rapp, ``Nearly 4 million bitcoins lost forever, new study
  says,'' 2017, accessed: 9/2/2018.

\bibitem{w3c2018issue}
M.~Sporny. \url{https://github.com/w3c/vc-data-model/issues/112}. Accessed:
  21/2/2018.

\bibitem{erc780}
J.~Torstensson, ``Ethereum claims registry,'' \emph{ERC 780}.

\bibitem{zug}
C.~of~Zug. \url{http://www.stadtzug.ch/de/bevoelkerung/dienste/digitaleid/
  ?action=showthema\&themenbereich\_id=1587\&thema\_id=5295}. Accessed:
  14/2/2018.

\bibitem{olympics}
N.~Benes. Announcing gnosis olympia.
  \url{https://blog.gnosis.pm/announcing-gnosis-olympia-5fb7e16dd259}.
  Accessed: 14/2/2018.

\bibitem{benet2014ipfs}
J.~Benet, ``Ipfs-content addressed, versioned, p2p file system,'' \emph{arXiv
  preprint arXiv:1407.3561}, 2014.

\end{thebibliography}

\end{document}